\documentclass[a4paper,preprint,unsortedaddress]{revtex4-1}
\usepackage{amsmath,amssymb,amsfonts,latexsym}
\usepackage[dvips]{graphicx}
\usepackage{color}
\usepackage{indentfirst}

\newcommand{\recheck}[1]{{\color{black} #1}}

\newcommand{\vect}[1]{\textbf{\textit{#1}}}
\newcommand{\dof}{{\textrm{DOF}}}
\newcommand{\AT}{{\textrm{{AT}}}}

\newcommand{\CG}{{\textrm{CG}}}
\newcommand{\HY}{{\Delta}}

\newcommand{\thf}{{\textrm{th}}}
\newcommand{\res}{{\textrm{rep}}}
\newcommand{\ext}{{\textrm{extra}}}
\newcommand{\exc}{{\textrm{ex}}}
\newcommand{\thermo}{{\textrm{Q}}}
\newcommand{\hadress}{{\textrm{H}}}

\newcommand{\typea}{{\textrm{A}}}
\newcommand{\typeb}{{\textrm{B}}}
\newcommand{\concenttba}{x_{\textrm{TBA}}}

\begin{document}
\title{Chemical potential of liquids and mixtures via Adaptive Resolution Simulation}
\author{Animesh Agarwal}
\affiliation{Institute for Mathematics, Freie Universit\"at Berlin, Germany}
\author{Han Wang}
\email{han.wang@fu-berlin.de}
\affiliation{Institute for Mathematics, Freie Universit\"at Berlin, Germany}
\author{Christof Sch\"{u}tte}
\affiliation{Institute for Mathematics, Freie Universit\"at Berlin, and Zuse Institute Berlin (ZIB), Berlin, Germany}
\author{Luigi Delle Site}
\email{dellesite@fu-berlin.de}
\affiliation{Institute for Mathematics, Freie Universit\"at Berlin, Germany}

\begin{abstract}
We employ the adaptive resolution approach AdResS, in its recently developed Grand Canonical-like version (GC-AdResS) [Wang {\it et al.} Phys.Rev.X 3, 011018 (2013)], to calculate the excess chemical potential, $\mu^{ex}$, of various liquids and mixtures.
We compare our results with those obtained from full atomistic simulations using the technique of thermodynamic integration and show a satisfactory agreement. In GC-AdResS the procedure to calculate $\mu^{ex}$ corresponds to the process of standard initial equilibration of the system; this implies that, independently of the specific aim of the study, $\mu^{ex}$, for each molecular species, is automatically calculated every time a GC-AdResS simulation is performed. 
\end{abstract}

\maketitle

\section{Introduction}
The chemical potential represents an important thermodynamic information for any system, in particular for liquids, where the possibility of combining different substances for forming optimal mixtures is strictly related to knowledge of the chemical potential of each component in the mixture environment. In this perspective, molecular simulation represents a powerful tool for predicting the chemical potential of complex molecular systems. Popular, well established methodologies in Molecular Dynamics (MD) are Widom particle insertion (IPM) \cite{widom} and thermodynamic integration (TI) \cite{ti}. IPM is computationally very demanding often beyond a reasonable limit even in presence of large computational resources, but upon convergence, is rather accurate. TI is computationally convenient but specifically designed to calculate the chemical potential and thus  it may not be optimal for employing MD for studying other properties. In fact TI requires artificial modification of the atomistic interactions (see Appendix).
Recently we have suggested that the chemical potential could be calculated by employing the Adaptive Resolution Simulation method in its Grand Canonical-like formulation (GC-AdResS) \cite{prl12,jctchan,prx}.
AdResS was originally designed  to interface regions of space at different levels of molecular resolution within one simulation set up. This allows for large and efficient multiscale simulations where the high resolution region is restricted to a small portion of space and the rest of the system is at coarser level. The recent version of the method, GC-AdResS, given its theoretical framework, should automatically calculate the chemical potential during the process of initial equilibration: in this work we prove that this is indeed the case and report results for the chemical potential for various liquids and mixtures of particular relevance in (bio)-chemistry and material science. We compare our results with those from full atomistic TI and find a satisfactory agreement. This agreement allows us to conclude that every time a multiscale GC-AdResS is performed, $\mu^{ex}$ is automatically calculated for each liquid component and implicitly confirm that the basic thermodynamics of the system is well described by the method.
{Moreover, in recent work AdResS has been merged with the MARTINI force field \cite{matej-sw1,matej-sw2}. In this context, the possibility of checking the consistency of a quantity like the chemical potential can be used as a further argument for the validity of the method in applications to large systems of biological interest.}
Below we provide the basic technical ingredients of GC-AdResS which are relevant for the calculation of the chemical potential, more specific details can be found in~\cite{jctchan, prx}.

\section{From AdResS to GC-AdResS}
\begin{figure}
\center
\includegraphics[width=0.475\textwidth]{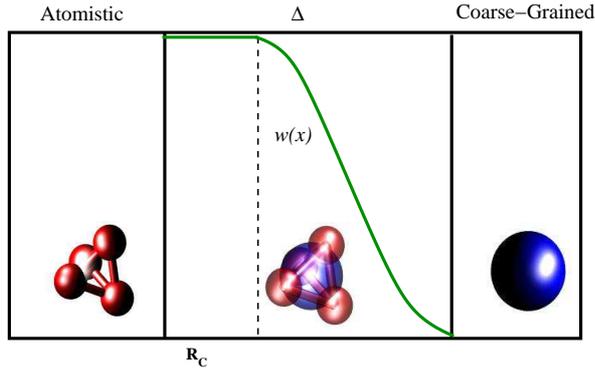}
\caption{Pictorial representation of the adaptive box and
molecular representation. Here it is shown the case of the tetrahedral molecule used as a test case in the original development of AdResS. The region on the right,
is the low resolution region (coarse-grained), the central part is
the transition (hybrid) region $\Delta$, where the switching function
$w(x)$ (in green) is defined, and the region on the left, is the high resolution region (atomistic). It must be noticed that differently from the original AdResS, in GC-AdResS the range of definition of $w(x)$ is extended of an amount of ${\vect R}_{c}$. The extension of this additional region is equal to the cut-off radius of the atomistic interactions and $w(x)$ takes the constant value of $1$. The consequence is that molecules in the atomistic region interact with the rest of the system always and only via well defined atomistic interactions. This characteristic, in turn, allows to write an exact Hamiltonian for the atomistic region and thus treat the system in a Grand-Canonical fashion {(see Eq.12 of Ref.\cite{prx})}}.
\label{fig1}
\end{figure}

The original idea of AdResS is based on a simple intuitive physical principle:
\begin{itemize}
\item Divide the space in three regions, one with atomistic resolution (AT) and one with coarse-grained (spherical) resolution (CG) interfaced by a smaller region with an hybrid treatment, which is usually called transition region or hybrid region.
\item Couple the molecules in the different regions through a spatial interpolation formula on the forces:
\begin{equation}
{\vect F}_{i,j}=w(\vect r_i)w(\vect r_j){\vect
  F}_{i,j}^{\AT}+[1-w(\vect r_i)w(\vect r_j)]{\vect F}^{\CG}_{i,j} 
\label{eqforce}
\end{equation}
where $i$ and $j$ indicates two molecules, ${\vect F}^{\AT}$ is the
force derived from the atomistic force field and  ${\vect F}^{\CG}$
from the corresponding coarse-grained potential, {$\vect r$ is the center of mass (COM) position of}
the molecule and $w$ is an interpolating function which smoothly goes from $0$
to $1$ (or vice versa) in the transition region ($\Delta$) where the lower resolution is then
slowly transformed (according to $w$) in the high resolution (or vice versa),
as illustrated in Fig.\ref{fig1}.
\item In the transition region a thermodynamic force acting on the COM of each molecule and a locally acting thermostat are added to assure the overall thermodynamic equilibrium at a given temperature. The thermodynamic force is defined in such a way that
  $p_{\AT}+\rho_{0}\int_{\Delta}{\vect F}_{\thf}({\vect r})d{\vect r}=p_{\CG}$,
  where $p_{\AT}$  is the target pressure of the atomistic system (region), $p_{\CG}$ is the pressure of the coarse-grained model, $\rho_{0}$ is the target molecular density of the atomistic system (region)~\cite{prl12}. An additional locally acting thermostat is added to take care of the loss/gain of energy in the transition region.
\end{itemize}
In \cite{prx} we have defined necessary conditions in $\Delta$ such that the spatial probability distribution of the full-atomistic reference system was reproduced up to a certain (desired) order in the atomistic region of the adaptive system. {We have defined the $m$th order of a spatial (configurational) probability distribution of $N$ molecules, $p({\vect r}_{1}, \cdots, {\vect r}_{N})$, as:
\begin{equation}
p^{(m)}({\vect r}_{1}, \cdots, {\vect r}_{m})=\int p({\vect r}_{1}, \cdots, {\vect r}_{m},{\vect r}_{m+1}, \cdots, {\vect r}_{N}) \: d{\vect r}_{m+1}\cdots d{\vect r}_{N}
\label{order}
\end{equation}
The first order, often mentioned in this work, corresponds to the molecular number density $\rho({\vect r})$.}
Moreover we have shown that, because of the necessary conditions, the accuracy in the atomistic region is independent of the accuracy of the coarse-grained model, thus, in the coarse-grained region, one can use a generic liquid of spheres whose only requirement is that it has the same molecular density of the reference system. In the simulation set up, ${\vect F}_{th}$ is calculated via an iterative procedure {using} the molecular number density in $\Delta$. The iterative scheme consists of calculating {${\vect F}^{k+1}_{th}({\vect r}) = {\vect F}^{k}_{th}({\vect r}) + \frac{1}{\rho^2_{0}\kappa_{T}}\nabla \rho^{k}({\vect r})$} ($\kappa_{T}$ the isothermal compressibility), and the thermodynamic force is considered converged when the target density $\rho_{0}$ is reached in $\Delta$. As a result, ${\vect F}_{th}({\vect r})$, acting in $\Delta$, assures that there are no artificial density variations across the system, thus it allows to accurately reproduce the first order of the probability distribution in the atomistic region. Higher orders can be systematically achieved by imposing in $\Delta$ a corrective force.
{For example, the COM-COM radial distribution function correction for the second order~\cite{jctchan}.} Next it was proved that indeed the target Grand Canonical distribution, that is the probability distribution of a subsystem (of the size of the atomistic region in GC-AdResS) in a large full atomistic simulation is accurately reproduced. A large number of tests were performed and the reproduction by GC-AdResS of the probability distribution was numerically proved up to (at least) the third order, more than sufficient in MD simulations. Within this framework it was finally shown that the sum of work of ${\vect F}_{th}({\vect r})$ and that of the thermostat corresponds to the difference in chemical potential between the atomistic and coarse-grained resolution; this subject is treated in the next section.

\section{Calculation of Excess Chemical Potential}

In Ref.~\cite{prx} it has been shown that the chemical potential of the atomistic and coarse-grained resolution are related by the following formula: 
\begin{equation}
\mu_{\CG}=\mu_{\AT}+\omega_{\thf}+\omega_{\thermo}
\label{mu}
\end{equation}
with $\mu_{\CG}$ the chemical potential of the coarse-grained system (in GC-AdResS this corresponds to a liquid of generic spheres), $\mu_{\AT}$  the chemical potential of the atomistic system, $\omega_{\thf}=\int_{\Delta}{\vect F}_{\thf}({\vect r})d{\vect r}$ the work of the thermodynamic force in the transition region, $\omega_{\thermo}$ the work/heat provided by the thermostat in order to slowly equilibrate the inserted/removed degrees of freedom in the transition region. $\omega_{\thermo}$ is composed by two parts, one, called $\omega_{\ext}$, which compensates the dissipation of energy due to the non-conservative interactions in $\Delta$, and another, $\omega_{\dof}$, which is related to the equilibration of the reinserted/removed degrees of freedom (rotational and vibrational). {While the determination of $\omega_{\dof}$ is not required for our final aim (that is the calculation of the excess chemical potential, as explained later on), the calculation of $\omega_{\ext}$ is very relevant}. However this calculation is not straightforward and we have proposed to introduce an auxiliary Hamiltonian approach where the coarse-grained and atomistic potential are interpolated, and not the forces as in {the original} AdResS. Next, we impose that the Hamiltonian system must have the same thermodynamic equilibrium of the original force-based GC-AdResS system; this is done by introducing a thermodynamic force in the auxiliary Hamiltonian approach, which, at the target temperature, keeps the density of particles across the system as in GC-AdResS. {In the auxiliary Hamiltonian approach we have the same equilibrium as the original adaptive (and full atomistic) system  and the difference between the work of the original thermodynamic force and the work of the thermodynamic force calculated in the Hamiltonian approach gives $\omega_{\ext}$ (further details about this point are given in the {Appendix~\ref{app:tmp2}})}. Moreover we have proven numerically, for the case of liquid water, that
  $\omega_{\ext}=\int_{\Delta}\nabla w(\vect r)\langle w( U^{\AT}-U^{\CG})\rangle_{\vect r} d{\vect r}$,
  where $U_{\AT}$ and $U_{\CG}$ are the atomistic and coarse-grained potential.
{It must be noticed that the auxiliary Hamiltonian approach shall not be considered a Hamiltonian approach to adaptive resolution simulation. In fact, as discussed in Ref.\cite{prx} the equilibrium is imposed artificially and {\it per se} does not have any physical meaning (for more details, see discussion in the {Appendix~\ref{app:tmp2}})}. 
{In the next section of this work we show analytically that the formula above is exact (at least) at the first order w.r.t. the probability distribution of the system as defined in Eq.\ref{order}}.
The result above implies that $\omega_{\thermo}$ can be calculated in a straightforward way during the initial equilibration within in the standard GC-AdResS code. 
{It must be noticed that, within the AdResS scheme, an approach similar to the auxiliary Hamiltonian  has been recently proposed and applied to liquids and mixtures (of toy models so far) by Potestio {\it et al.} \cite{h-adress-0, h-adress} (see also \cite{luigientropy} where such an approach is commented)}. At this point according to \eqref{mu}, if one knows $\mu_{\CG}$, then  GC-AdResS can automatically provide $\mu_{\AT}$. However we need to do one step more, in fact the quantity of interest is not the total chemical potential, but the excess chemical potential $\mu^{\exc}_{\AT}$ which corresponds to the expression of \eqref{mu} where the kinetic (ideal gas) part is subtracted. Regarding the kinetic part, one can notice that the contribution coming from the COM is the same for the coarse-grained and for the atomistic molecules, thus it is automatically removed in the calculation of \eqref{mu}.
The kinetic part of $\mu_{\AT}$ due to the rotational and vibrational degrees of freedom corresponds in our case to $\omega_{\dof}$ and {in principle can be calculated by hand (chemical potential of an isolated molecules). However such a calculation may become rather tedious for large and/or complex molecules but in our case it is actually not required.} In fact the {Gromacs implementation of AdResS} considers the removed degrees of freedom as phantom variables but thermally equilibrate them anyway~\cite{simon-ch}. Thus the heat provided by the thermostat for the rotational and vibrational part is the same in the atomistic and coarse-graining molecules and is automatically removed in the difference. Finally, the calculation of $\mu^{\exc}_{\CG}$ can be done with standard methods, TI or IPM, which for simple spherical molecules, like those of the coarse-grained system, requires a negligible computational cost.
In conclusion, we have the final expression:
\begin{equation}
  \mu^{\exc}_{\AT}=
  \mu^{\exc}_{\CG}-
  \int_{\Delta}{F}_{th}({\vect r})d{\vect r}-
  \int_{\Delta}\nabla_{\vect r} w(\vect r) \langle w(U^{\AT}-U^{\CG})\rangle_{\vect r} d{\vect r}
\label{finalmu}
\end{equation}

\section{Analytic Derivation of  $\omega_{\ext}$}\label{sec:tmp4}
{In this section we derive analytically the equivalence:
  $\omega_{\ext}=\int_{\Delta}\nabla_{\vect r} w(\vect r) \langle w(U^{\AT}-U^{\CG})\rangle_{\vect r} d{\vect r}$ and define its conceptual limitations.
We consider a potential coupling between the atomistic and coarse-grained resolution, that is the spatial interpolation of the atomistic and coarse-grained potential, as done instead for the forces in the standard AdResS:
\begin{align}
  U =
  \sum_{i<j} w(\vect r_i) w(\vect r_j)
  U^\AT_{i,j}
  +
  \sum_{i<j}[1- w(\vect r_i) w(\vect r_j)]
  U^\CG_{i,j},
\end{align}
where $U^\AT_{i,j}$ and $U^\CG_{i,j}$ are the atomistic and coarse-grained interaction potential between molecule $i$ and $j$, respectively, defined by
\begin{align}\label{eqn:pot-interpol}
  U^\AT_{i,j} =
  \sum_{\alpha\in i}\sum_{\beta\in j} U^\AT(\vect r_\alpha - \vect r_\beta),\quad
  U^\CG_{i,j} =
  U^\CG(\vect r_i - \vect r_j),
\end{align}
where  $\alpha$ and $\beta$ denotes the atom indices of the corresponding molecule.
The COM of the molecule is defined as:
\begin{align}
  \vect r_i = \sum_{\alpha\in i} \frac{m_\alpha}{\sum_{\alpha\in i} m_\alpha} \vect r_\alpha,
\end{align}
where $m_\alpha$ is the mass of atom $\alpha$ of molecule $i$.
{The potential interpolation~\eqref{eqn:pot-interpol} provides an auxiliary Hamiltonian to the AdResS system, and
the
corresponding intermolecular force is given by:}
\begin{align}\label{eqn:interpol-f}
  \vect F_{i,j} =
  w(\vect r_i) w(\vect r_j)
  \vect F^\AT_{i,j}
  +
  [1- w(\vect r_i) w(\vect r_j)]
  \vect F_{i,j}^\CG
  -
  \nabla_{\vect r} w(\vect r_i) w(\vect r_j)
  (U^\AT_{i,j} - U^\CG_{i,j}).
\end{align}
{We refer to the AdResS simulation using potential scheme~\eqref{eqn:interpol-f} as
  auxiliary Hamiltonian AdResS, and all properties of this approach will be added a superscript ``H''}.
We define the force of changing representation by
\begin{align}\label{eqn:frep}
  \vect F_{\res,i} = 
  \sum_j \nabla_{\vect r} w(\vect r_i) w(\vect r_j)
  (U^\AT_{i,j} - U^\CG_{i,j}).
\end{align}

We use the same notation as in our previous work~\cite{prx}.
The thermodynamic variables for the atomistic and coarse-grained regions
are denoted by $(N_1, V_1, T)$ and $(N_3, V_3, T)$, respectively.
We assume that the transition region is an infinitely thin filter (that is a much smaller region than the atomistic and coarse-grained region) that allows molecules
to change resolution as they cross it. Therefore, it is 
reasonable to assume that:
\begin{align}
  V &= V_1 + V_3\\
  N &= N_1 + N_3
\end{align}
where $V$ and $N$ are the total volume and total number of
molecules of the system. In this work, we adopt the same
assumptions as those listed in Sec.III.C of Ref.~\cite{prx},
i.e. we assume the system to be in the thermodynamic limit, and molecules are short-range
correlated (short-ranged must be intended as a range comparable to the size of the transition region). The thermodynamic force for GC-AdResS ($\vect F_\thf$) and for the auxiliary Hamiltonian AdResS ($\vect F^\hadress_\thf$),
enforce the system to have a flat density:
\begin{align}
  \rho_\HY &= \rho_\AT = \rho_\CG = \rho_0\\
  \rho_\HY^\hadress &= \rho_\AT = \rho_\CG = \rho_0
\end{align}
Where $\rho_0$ is the equilibrium number density of the system defined by $\rho_0 = N/V$.
As shown in Refs.\cite{prl12, prx}, $\vect F_\thf$ provides the
balance of the grand potential or equivalently
\begin{align}\label{eqn:peq}
  p_\AT = p_\CG - \rho_0 \,\omega_\thf 
\end{align}
where $ \omega_\thf$ denotes the work of the thermodynamic force $\vect F_\thf$. \\

Instead when we consider the auxiliary Hamiltonian approach, the third term on the R.H.S.~of Eq.~\eqref{eqn:interpol-f} is not
symmetric w.r.t molecules $i$ and $j$, therefore, the Newton's
action-reaction law (momentum conservation) does not hold anymore.
As a consequence, the pressure relation between the AT and CG
resolution~\eqref{eqn:peq} does not hold and should be derived again.
Now assume, {for simplicity and without lost of generality}, that the system changes resolution only along the $x$ direction.
We impose an infinitesimal increment of the volume $\Delta V$ to the
AT region, and apply the same decrement of the volume $-\Delta V$ to the CG
region.  The volume of the transition region is kept constant as if it is an ideal ``piston'' that moves toward the CG region by an amount $\Delta L$.
We assume $\Delta V = \Delta L\cdot S$, where $S$ is the
cutting surface area.
The displacement $\Delta L$ should be infinitesimal, i.e.~much smaller than the size of the
transition region. This is achievable by taking the limit of $\Delta L\rightarrow 0$,
while keeping the system size fixed.
It must be noticed that also the displacements of the molecules are infinitesimal, so it can be reasonably assumed that
the resolution of the molecules remains the same under a displacement of $\Delta L$.
Therefore, the change of the free energy of the system is approximately:
\begin{align}\nonumber
  \Delta A \approx&\,
  A_\AT(N_1, V_1+\Delta V, T) -
  A_\AT(N_1, V_1, T)
  +
  A_\CG(N_3, V_3-\Delta V, T) -
  A_\CG(N_3, V_3, T)\\\label{eqn:apptmp3}
  &-
  \recheck{\int_{\Delta_{x}} d x\, \rho_0 S\cdot \Delta L\cdot
  \big[\vect F^\hadress_\thf(x)
  -
  \big\langle \vect F_\res(x)\big\rangle\,\big]}
\end{align}
where $A_\AT$ and $A_\CG$ are the free energies of the AT and CG region, respectively.
\recheck{
$\Delta_{x}$ is the linear dimension of the transition region along $x$. 
Since the resolution changes only along $x$,
the two one-particle forces depend only on $x$, and only have the component along $x$.
This can be easily generalized to changing resolution in any direction, i.e., replacing $x$ by ${\vect r}$.}
The expression of Eq.\ref{eqn:apptmp3} as a sum of different terms is justified by the hypothesis of treating the system in the thermodynamic limit, and by the hypothesis that the interactions are short-ranged compared to the size of the transition region.
$N_1$ and $N_3$ is the numbers of molecules in the AT and CG region, and
$V_1$ and $V_3$ is the volume in the AT and CG region, respectively; $T$ is the temperature
of the system.
The last term is originated by the work done by the ideal piston. This term is composed by
two parts, the first corresponding to the work done by the thermodynamic force, and the
second corresponding to 
the work done by the force of changing representation (which does not vanish due to the violation of the Newton's
action-reaction law). The first and second term of Eq.~\eqref{eqn:interpol-f}
being forces based on pairwise interactions only, do not contribute to the difference of energy; in fact 
their total work is zero (as long as the transition region moves infinitesimally along $x$).
The notation $\langle\cdot\rangle$ in Eq.~\eqref{eqn:apptmp3} denotes the ensemble average, which
will be specified soon.
It is straightforward to show that
\begin{align}\label{eqn:peq-d}
  \Delta A \approx
  -p_\AT\Delta V + p_\CG\Delta V -
  \rho_0 \Delta V (\omega_\thf^\hadress - \omega_\res), 
\end{align}
where $\omega_\thf^\hadress$ is the work of the 
thermodynamic force $\vect F_\thf^\hadress$, and
$ \omega_\res$ is the work of changing representation,
which can be explicitly written down \recheck{in a general form} as:
\begin{align}\label{eqn:w-rep}
  \omega_\res
  = \int_\HY d \vect r\,\langle \vect F_\res (\vect r)\rangle
  = \int_\HY d \vect r\,\nabla_{\vect r}w(\vect r)\,
  \big\langle w(\vect r')\,
  [\sum_{\alpha,\beta}U^\AT(\vect r_\alpha-\vect r'_\beta) - U^\CG(\vect r-\vect r')]\big\rangle_{\vect r'; \vect r},
\end{align}
The average is performed over all
  possible positions of the second molecule (i.e.~$\vect r'$),
  at fixed position of the first molecule (i.e.~$\vect r$) 
  in the pairwise interaction.
In case of molecules containing
more than one atom, the average is also made over all possible
conformations in the atomistic resolution.
In the thermodynamic limit, the equilibrium volume of the AT region maximizes the
free energy, i.e.~$\Delta A / \Delta V
= 0$, which yields
\begin{align}\label{eqn:peq-h}
  p_\CG - p_\AT =  \rho_0 (\omega_\thf^\hadress - \omega_\res).
\end{align}
Comparing the expression above with that obtained for GC-AdResS~(Eq.~\eqref{eqn:peq}),
we have: 
\begin{align}\label{eqn:hd-rel}
  \omega_\res = \omega_\thf^\hadress -   \omega_\thf,
\end{align}
which relates the thermodynamic force of the auxiliary Hamiltonian AdResS
and the GC-AdResS.\\

In Ref.~\cite{prx} we proved that under proper assumptions,
when the flat density profile is enforced by the thermodynamic force,
the chemical potential difference between the different resolutions
is given by
\begin{align}\label{eqn:mueq-d}
 \mu_\CG - \mu_\AT =   \omega_\thf + \omega_\dof + \omega_\ext
\end{align}
The same argument can be applied to the auxiliary Hamiltonian approach,
and yields  the chemical potential difference between the AT and CG resolutions
\begin{align}\label{eqn:mueq-h}
 \mu_\CG - \mu_\AT =   \omega^\hadress_\thf + \omega_\dof 
\end{align}
In the auxiliary Hamiltonian, we do not have the term $\omega_\ext$ in the above
formula {(being the term $\omega_\ext$ in GC-AdResS, generated by the non-conservative effect of the force interpolation).}
By comparing~\eqref{eqn:mueq-d} with~\eqref{eqn:mueq-h}, we have the
relation
\begin{align}\label{eqn:hd-rel-2}
  \omega_\ext = \omega^\hadress_\thf - \omega_\thf,
\end{align}
which also relates the thermodynamic force of the auxiliary Hamiltonian AdResS
and GC-AdResS.\\

From Eq.~\eqref{eqn:hd-rel} and \eqref{eqn:hd-rel-2}, we
find the extra work of the thermostat in GC-AdResS being identical to the work of changing
representation of the auxiliary Hamiltonian approach:
\begin{align}\label{eqn:wextra}
  \omega_\ext = \omega_\res,
\end{align}
which basically proves the statement at the beginning of this section.  The ensemble
average on the R.H.S.~of Eq.~\eqref{eqn:w-rep} is performed in the
ensemble of the system treated with the potential interpolation approach, and the question is if the ensemble average is equivalent if it is performed in the simulation where the force
interpolation approach is used.  It is obvious that the spatial probability distribution corresponding to the system treated with the potential interpolation is consistent with the force
interpolation at least up to the first order.  It is also
possible to systematically obtain equivalence in the ensemble average operation  at higher orders of accuracy of the probability distribution, as, for example, it is done for the radial distribution function in Ref.~\cite{jctchan}.
However, here we do not consider higher order corrections, because it
has been numerically shown that actually the ensemble average of
$\vect F_\res$ dose not depend on in which ensemble it is
calculated~\cite{prx}. Therefore, we use Eq.~\eqref{eqn:wextra} to
calculate $\omega_\ext$, and measure the ensemble average by the
standard AdResS.  As previously discussed, in the Gromacs implementation, the CG
molecules also keep the atomistic degrees of freedom even though they
are in the CG region, therefore, the kinetic part of $\mu_\AT$ and
$\mu_\CG$ are identical, and $\omega_\dof$ vanishes. Therefore,
by inserting Eq.~\eqref{eqn:wextra} into \eqref{eqn:mueq-d}, we have
\begin{align}\label{eqn:tmp24}
  \mu_\AT^\exc = \mu_\CG^\exc - \int_\HY d\vect r \vect F_\thf(\vect r)
-\int_\HY d \vect r\,\nabla_{\vect r}w(\vect r)\,
\big\langle w(\vect r')\, [\sum_{\alpha,\beta}U^\AT(\vect r_\alpha-\vect r'_\beta) - U^\CG(\vect r-\vect r')]\big\rangle_{\vect r'; \vect r}
\end{align}
The extension of Eq.\ref{eqn:tmp24} to multicomponent systems is
reported in the Appendix~\ref{app:tmp3}, while in the next section we
apply the method to the calculation of $\mu^{ex}$ to liquids and
mixtures.}

\section{Results and Discussion}
We have calculated $\mu^{ex}$ for different liquids and mixtures, choosing cases which are representative of a large class of systems. Hydrophobic solvation in methane/water and in ethane/water mixtures, hydrophilic solvation in urea/water, a balance of both in water/tert-Butyl alcohol (TBA) mixture, other liquids, e.g. pure methanol and DMSO (and their mixtures with water), non aqueous mixtures in TBA/DMSO and alkane liquids such as methane, ethane and propane. Moreover, systems as water/urea are commonly used as cosolvent of biological molecules \cite{nico-debashish} while systems as tert-Butyl alcohol/water play a key role in modern technology \cite{irata}, thus they are of high interest {\it per se}. All technical details of each simulation are presented in the Appendix~\ref{app:tmp1}.\\

Results are reported in Table \ref{table}, where the comparison with values obtained using full atomistic TI and available experiments, at the same concentrations, of our calculation is made; in our previous work we have already shown that value of the chemical potential of liquid water obtained with IPM is well reproduced by GC-AdResS, however the computational cost of IMP was very large, thus we do not consider calculations done with IPM in this paper. 
The agreement with full atomistic TI simulations is satisfactory in all cases, and thus it proves the solidity of GC-AdResS in describing the essential thermodynamics of a large class of systems.
We also compare the obtained values with those available in literature \cite{vang,nico}. Although the concentration of the minor component in the mixtures that we consider, is higher than the concentrations considered in Refs.\cite{vang,nico}, we are anyway in the very dilute regime and thus the chemical potential should not change in a significant way; {we have verified  such a supposed consistency for one relevant system (see discussion about Fig.\ref{concentration})}. 
The chemical potential of $k$-th liquid's component in a mixture is calculated as {(see Appendix~\ref{app:tmp3})}:
\begin{equation}
  \mu^{\exc,k}_{\AT}=
  \mu^{\exc,k}_{\CG}-
  \int_{\Delta}{\vect F}^k_{\thf}({\vect r})d{\vect r}-
  \int_{\Delta}\nabla_{\vect r} w(\vect r) \langle w(U^{\AT}-U^{\CG})\rangle_{\vect r, k} d{\vect r}
\label{eqn:tmp25}
\end{equation}
where $\vect{F}^{k}_{\thf}(x)$ is the thermodynamic force applied to the molecules of the $k$-th component; this assures that, at the given concentration, the density of molecules of species $k$, in the transition region, is equivalent to the density of the same liquid's component in a reference full atomistic simulation.
{The ensemble average is taken over the position of the second molecule, provided that the first molecule is of species $k$, and located at position $\vect r$}.\\

\begin{table}[]
\begin{center}
\begin{tabular}{ccrrc}
\hline \hline
 Liquid component & Mole fraction of solute & GC-AdResS & TI & Experiment \\
\hline
{water} & -- & $-22.8 \pm 0.2$  & $-22.1 \pm 0.3$ & $-23.5$~\cite{florian} \\
methane  & -- & $-4.6 \pm 0.1$  & $-5.2 \pm 0.1$ & -- \\
ethane   & -- & $-8.2 \pm 0.3$  & $-8.8 \pm 0.1$ & -- \\
propane  & -- & $-8.5 \pm 0.1$ & $-9.5 \pm 0.2$ & -- \\
{methanol}  & -- & $-20.1 \pm 0.1$ & $-20.6 \pm 0.4$ & $-20.5$ \cite{vang}  \\
DMSO & -- & $-32.2 \pm 0.3$ & $-34.7 \pm 0.7$ & $-32.2$ \cite{dmso}  \\
{methanol in methanol/water mixture} & 0.01 & $-18.1 \pm 0.2$ & $-19.7 \pm 0.2$ & -- \\
{methane in methane/water mixture} & 0.006 & $9.1 \pm 0.1$  & $8.5 \pm 0.2$ & -- \\
urea in urea/water mixture & 0.02 & $-56.1 \pm 0.6$ & $-58.2 \pm 0.5$ & $-57.8 \pm 2.5$ \cite{urea} \\
{ethane in ethane/water mixture} & 0.006 & $7.2 \pm 0.2$ & $7.4 \pm 0.3$ & -- \\ 
{TBA in water/TBA mixture} & 0.001 & $-19.5 \pm 0.3$ & $-20.8 \pm 0.6$ & $-19.0$ \cite{nico} \\
DMSO in DMSO/water mixture & 0.01 & $-31.4 \pm 0.5$ & $-33.2 \pm 0.3$ & -- \\
TBA in TBA/DMSO mixture & 0.02 & $-24.8 \pm 0.4$ & $-24.0 \pm 0.5$ & -- \\
\hline \hline
\end{tabular}
\caption{The excess chemical potential of different liquids and mixtures in kJ/mol calculated from GC-AdResS and TI of full atomistic simulations. Experimental values for systems at the same concentrations used in simulation are also reported for comparison. For pure systems (water and methanol) we compare our values with those obtained in literature using the same force field and computational code. For mixtures, most of the values from literature (simulation and experiments) are available at lower concentrations (see Refs.\cite{vang,nico}); However, since we are always in a very dilute regime the chemical potential does not change significantly. {We have provided evidence for the TBA/water mixture that such consistency holds (see Fig.\ref{concentration})}. Note that the chemical potential of water in dilute mixtures is the same of pure water and is not reported above.
}
\label{table}
\end{center}
\end{table}

\begin{figure}
\center
\includegraphics[width=0.475\textwidth]{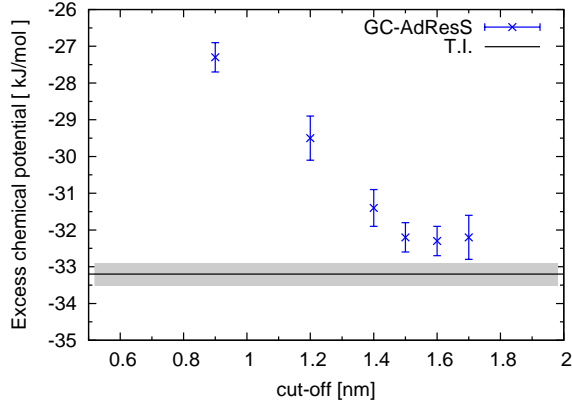}
\caption{ {Excess chemical potential of Dimethyl sulfoxide (DMSO) in water as a function of cut-off radius calculated using GC-AdResS. The value obtained from thermodynamic integration calculation in also shown, with a gray region indicating the standard deviation. This value was calculated using a cut-off radius of $1.4$~nm. It was seen that the value does not change significantly if the cut-off radius is varied.}}
\label{convergence}
\end{figure}

\begin{figure}
\center
\includegraphics[width=0.475\textwidth]{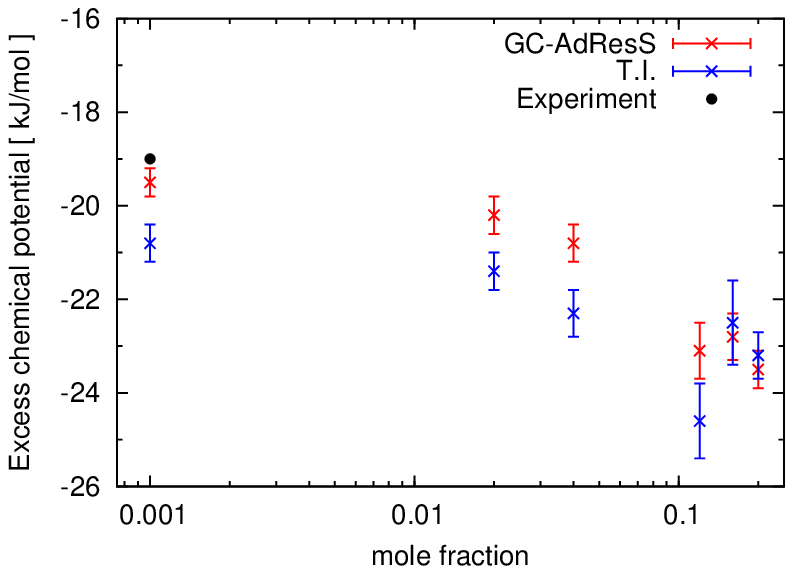}
\caption{ {Excess chemical potential of tert-butyl alcohol (TBA) in water for different concentrations (in logarithmic scale), calculated using GC-AdResS. The results are compared with thermodynamic integration values. At mole-fraction $\concenttba=0.001$, the experiment value is shown. \label{concentration}}}
\end{figure}

{A complementary information to Table~\ref{table} are 
  Fig.~\ref{convergence} and~\ref{concentration}.
  In Fig.~\ref{convergence} we have studied the behavior of $\mu^{ex}$ as a function of
  the interaction cut-off. In fact the current version of GC-AdResS,
  employs the reaction field method for treating electrostatic
  interactions in the atomistic region, and the cut off is likely to
  play a role in some of the systems investigated. Fig.\ref{convergence}, for the case of DMSO/water
  mixture, confirms our intuition and suggests that we could
  systematically improve the accuracy by increasing the cut-off,
  and at a value of about $1.5$~nm, $\mu^{ex}$ converges. In any
  case, at a values of $1.4$~nm, which is the one routinely used in
  full atomistic simulations and used by us, the
  value obtained with GC-AdResS is already satisfactory.
  The cut-off radii used for other systems are reported in Appendix~\ref{app:tmp1}.
  A further
  question that may arise is the capability of our method to predict
  the behavior of $\mu^{ex}$ as a function of the concentration, above
  all in the very dilute regime. In Fig.\ref{concentration} we have
  performed such a study for the case of TBA/water mixture, we show a
  good agreement between GC-AdResS and TI and show that at a very
  dilute concentration our calculated value is close to that of
  experiments, moreover the trend, regarding the TI calculations, is
  consistent with that reported in Ref.\cite{nico}.\\
}

\begin{figure}
\center
\includegraphics[width=0.475\textwidth]{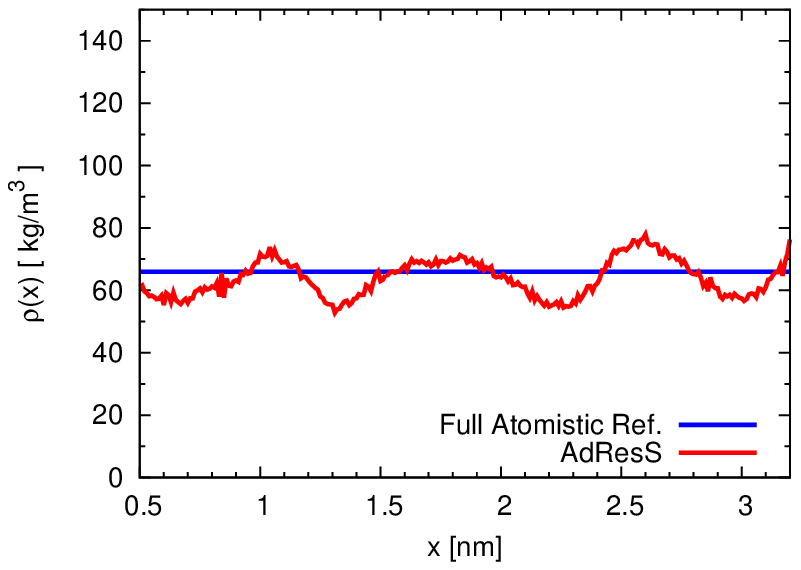}\\
\includegraphics[width=0.475\textwidth]{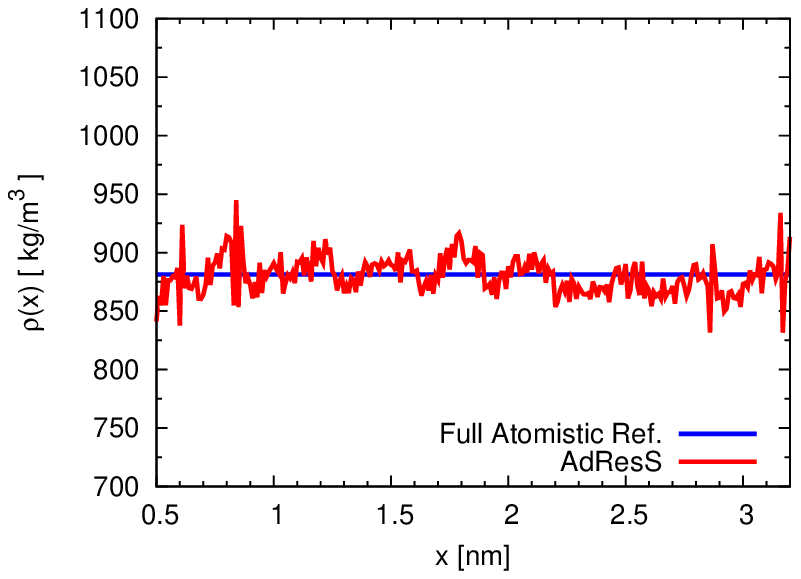}
\caption{Top: Molecular density profile in $\Delta$ for TBA/water mixture; Bottom, the same plot for water.
  Among all the systems considered, in this case the action of the thermodynamic force and that of the thermostat leads to the largest deviation from the reference all atomistic average density;
  however even in this case the discrepancy is negligible. The mole-fraction is $\concenttba=0.02$, and the cut-off radius is 0.9~nm.}
\label{urea-TBA}
\end{figure}

\begin{figure}
\center
\includegraphics[width=0.475\textwidth]{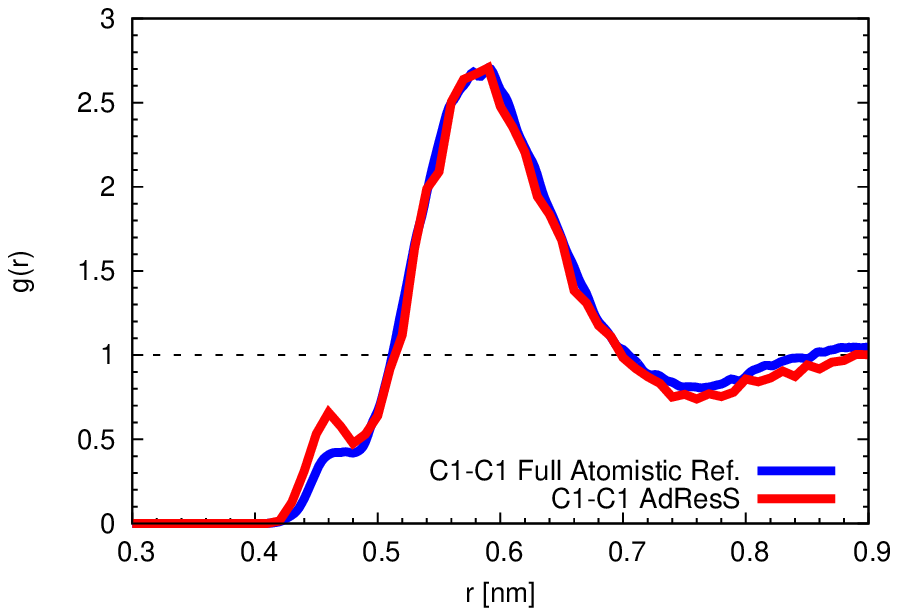}\\
\includegraphics[width=0.475\textwidth]{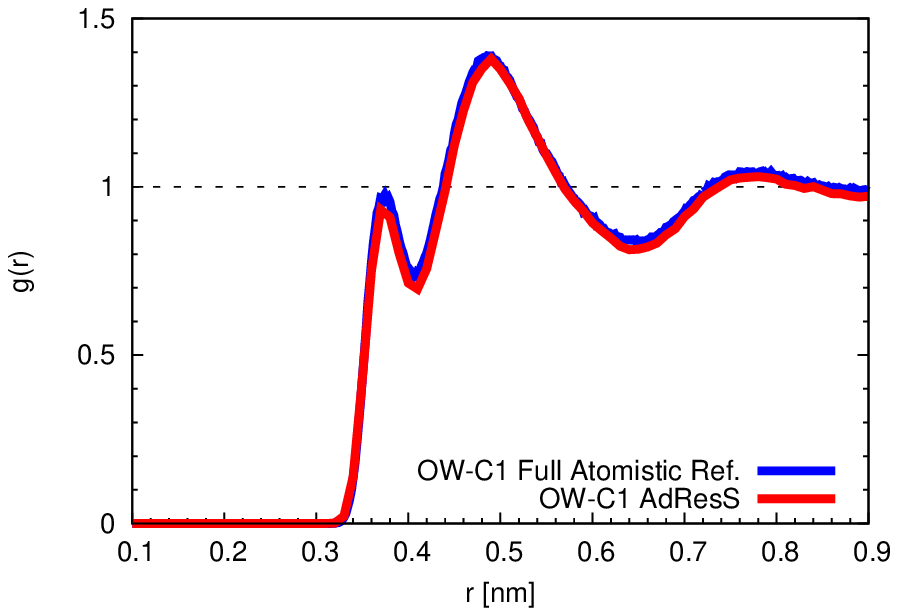}\\
\includegraphics[width=0.475\textwidth]{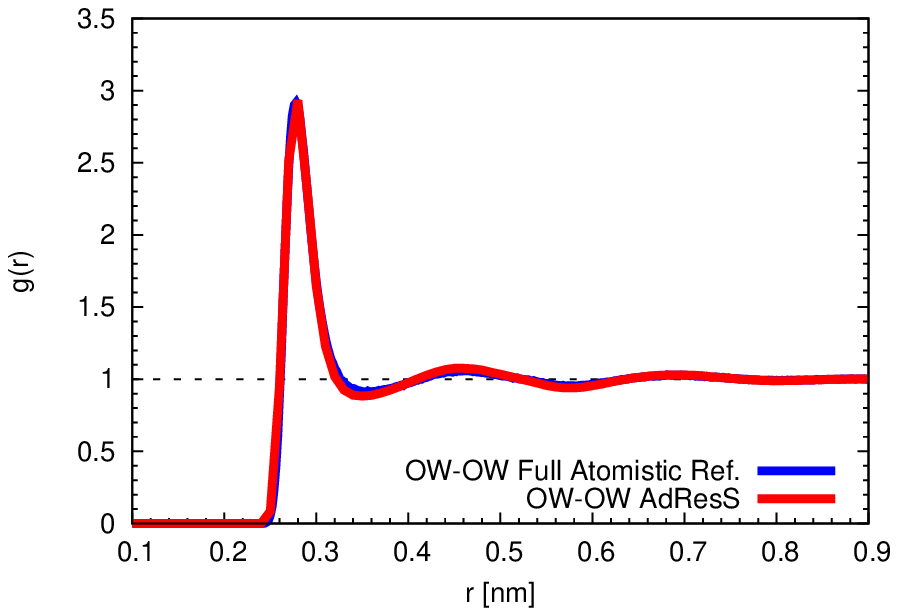}
\caption{Top: TBA-TBA radial distribution function; Middle: the same plot for TBA-water; Bottom: for water-water.
  Red: the results of the AdResS simulation.
  The $g(r)$ is calculated only in the atomistic region.
  Blue: the results of a full-atomistic reference simulation that has the same simulation region as AdResS.
  The mole-fraction is $\concenttba=0.02$, and the cut-off radius is 0.9~nm.}
\label{gr}
\end{figure}

In essence, according to the results obtained, GC-AdResS allows an on-the-fly determination of $\mu^{ex}$ of each component of a liquid, whenever a simulation is performed, without extra computational costs.
Moreover, Fig.\ref{urea-TBA} shows the action of the thermodynamic force and of the thermostat in the transition region $\Delta$ for TBA-water; the molecular density is sufficiently close to that of reference (the largest difference is below $20\%$ and the average difference is below $10\%$), and thus it assures that in the atomistic region there are no (significant) artificial effects on the molecular density due to the perturbation represented by the interpolation of forces in $\Delta$. In Fig.\ref{gr} we report various radial distribution functions for TBA-water in the atomistic region of the adaptive set up. The agreement with data from a full atomistic simulation is highly satisfactory. Moreover, it must be underlined that, on purpose, we have chosen extreme technical conditions, that is, a very small atomistic and coarse-grained region ($0.5$~nm) and a relatively large transition region ($2.7$~nm). Even in these conditions we prove that local properties as those of Fig.\ref{urea-TBA} and Fig.\ref{gr}, together with a relevant thermodynamics quantity as $\mu^{ex}$ are well reproduced.
This example shows the key features of GC-AdResS, that is, a multiscale simulation where the chemical potential of each component is obtained without extra computational costs and with high accuracy in a simulation where other properties are also calculated with high accuracy. It must be also noticed that the system corresponding to the figures is, among all the system considered, the case where the action of the thermodynamic force and of the thermostat produces the less accurate agreement with the reference data.

\section{Efficiency}

{In order to show the numerical efficiency of our approach, we compare the time taken to do a full GC-AdResS simulation and the time for a thermodynamic integration calculation for
different systems with varied concentration of TBA in water. The total time required for an GC-AdResS simulation consists of 
the time taken to obtain a converged thermodynamic force and the time taken to obtain the coarse-grained chemical potential. 
The time taken to complete TI procedure at each value of $\lambda$ is summed up to obtain the total time. In this work, the TI 
is done in two stages, first the van der Waals  interactions are coupled followed by the electrostatic coupling. At each stage, 21 
equally distributed values of $\lambda$ are used,
therefore, in total 42 simulations were performed to calculate each TI chemical potential value.
In an AdResS simulation, the initial guess of the thermodynamic force largely determines the time for convergence. 
We started with a randomly chosen initial guess
($-42$~kJ/mol, we picked a small value because the TBA molecule is hydrophilic)
for system with the highest mole fraction of TBA. For all the other systems, 
we used the converged thermodynamic force obtained from the first system as an initial guess. The convergence was much 
faster in all the other cases using this approach. Table~\ref{table3} shows the number of iterations required for 
the thermodynamic force convergence in GC-AdResS and total time required for GC-AdResS and TI calculation.
The advantage of GC-AdResS over TI
is that we get two values of excess chemical potential for both solute and solvent in a single calculation, while 
in TI, the whole process has to be repeated to get the excess chemical potential of the other component. For very dilute systems ($\concenttba=0.001$), 
however, one has to take a very large systems in GC-AdResS (see the Appendix~\ref{app:tmp1} for system size). It takes a large amount of time for the thermodynamic force to reach convergence, and hence TI is always a better option at such low concentrations with a much 
smaller system size at the same mole fraction. 

\begin{table}[htpb]
\begin{center}
\begin{tabular}{|l|c|r|r|}
\hline
 &\multicolumn{2}{c|}{GC-AdResS}&\multicolumn{1}{l|}{TI}\\
 \hline
$\concenttba$ &No. of iterations&Time(hrs.)&Time(hrs.)\\
\hline
0.200 & 20 & 52.4 & 30.8 \\
0.160 & 5 & 15.1 & 31.0 \\
0.120 & 5 & 16.3 & 29.9 \\
0.040 & 8 & 27.2 & 26.7 \\
0.020 & 8 & 28.8 & 26.6 \\  
0.001* & 20 & 252 & 202 \\
\hline
\end{tabular}
\caption{Time required for a full GC-AdResS and thermodynamic
  integration (TI) calculation for tert-butyl-alcohol(TBA) in water at
  different mole-fractions.
  The ``*'' means that a larger system was used for  the very dilute system ($\concenttba=0.001$), see Appendix~\ref{app:tmp1} for details.
  All the simulations were performed on a
  workstation that has two Quad-Core AMD Opteron(tm) 2376 Processors.
}
\label{table3}
\end{center}
\end{table}
}

\section{Current computational convenience: a critical appraisal}
The natural question arising from the discussion above is whether or not GC-AdResS is a more convenient technical tool for calculating $\mu^{ex}$ compared to TI.
Currently the answer is {neither negative nor positive}, although the current work is the first step towards a potentially positive answer for the future. In fact, the fastest version of AdResS is implemented in the GROMACS code \cite{gromacs}; using the Gromacs version 4.5.1 a speedup of a factor four with respected to full atomistic simulations has been reported for aqueous mixtures~\cite{debashish1,nico-debashish}. In this case GC-AdResS was more convenient than TI because in one simulation one could obtain the chemical potential of each liquid component and at the same time calculate structural properties (e.g. radial distribution functions). However in the successive version of GROMACS 4.6.1 the performance of atomistic simulations (above all of SPC/E water) has been highly improved while the corresponding implementation of AdResS is not optimized yet. At the current state, AdResS can only assure a speed up factor between {2 and 3} for large systems (30000 molecules) compared to full atomistic simulations {(except for pure SPC/E water systems)}. As a consequence for the calculation of $\mu^{ex}$, TI is in general computationally less demanding than AdResS . Another point that must be considered (in perspective) for a fair comparison between TI and GC-AdResS, is the following: even if AdResS is optimized, in any code, TI has the advantage that one can use one single molecule in the simulation box to mimic the minor component of a mixture. In our case, instead, we must treat, technically speaking, a true mixture with a certain number of molecules of the minor component immersed in the liquid of the major component. Thus, at low concentrations, GC-AdResS simulations require larger systems than those required by TI, moreover, because of the low density of the minor component, the convergence of the corresponding thermodynamic force requires long simulations. 
Thus, for very dilute systems, if one is interested only in the chemical potential, TI shall be preferred to GC-AdResS, however if the interest goes beyond the calculation of the chemical potential, (e.g. radial distribution functions) then (optimized) GC-AdResS would still be more convenient.
When the concentration becomes higher, GC-AdResS may become preferable for both tasks: general properties of the mixture and chemical potential, not only because in this case one requires larger systems, but also because the convergence of the thermodynamic force of the minor component is much faster. Moreover, we would have the flexibility of calculating the chemical potential
of both components in one simulation run, whereas in TI, one needs to run two separate simulations in order to get the chemical potential of both components.
{The results reported in the previous section about the current efficiency of GC-AdResS are rather encouraging, however currently there is not a clear convenience in using GC-AdResS instead of TI for calculating $\mu^{ex}$; 
in any case the technical aspects of code optimization must be reported and we must make clear that the aim of this work is to show that  the automatic calculation of $\mu^{ex}$, independently from the simulation code in which is implemented and its computational cost, is a ``conceptual'' feature of GC-AdResS}.
  
\section{Conclusion} 
We have shown the accuracy of GC-AdResS in calculating the excess chemical potential for a representative class of complex liquids and mixtures. 
For any system, the initial equilibration process, that is the determination of the thermodynamic force, automatically delivers the chemical potential. The only additional calculation required is that of $\mu^{ex}_{CG}$ which implies the use of IPM or TI, but for a liquid of simple spheres, thus computationally negligible.  
The essential message is that GC-AdResS would be, {\it per se}, a reliable multiscale technique to calculate the chemical potential and, in perspective, upon computational/technical optimization it may become an efficient tool for calculating $\mu^{ex}$ compared to current techniques in MD such as TI. 

\section*{Acknowledgments}
This work was supported by the Deutsche Forschungsgemeinschaft (DFG) with the Heisenberg grant provided to L.D.S (grant code DE 1140/5-1) and with its associate DFG grants for A.G.(grant code DE 1140/7-1) and for H.W. (grant code DE 1140/4-2).H.W.~and C.S.~thank the financial support by DFG research center MATHEON. {We thank Christoph Junghans and Debashish Mukherji for contributing to the information reported in section IV about the speed up factor}.

\appendix
\section{Technical details of the simulations}\label{app:tmp1}
The potential energy function and the force field parameters for all the molecules
were taken from GROMOS53A6 parameter set. Liquid water was described by the SPC
model~\cite{spc}, methanol was described by the model developed by Walser et al~\cite{walser},
urea by the model described in~\cite{urea}, tert-butyl alcohol by the parameter set of~\cite{nico}
and DMSO was described by the model given by Geerke et.al.~\cite{dmso1}. For liquid methanol simulations, 
GROMOS43A1 parameter set was used, as it was shown to be more accurate for calculating excess free energy of 
solvation of methanol in methanol~\cite{vang}. \\

In all the AdResS simulations, the resolution changes only along the $x$ direction.
For each system, 30 iterations were performed to obtain a 
converged thermodynamic force and a flat density profile. Each iteration consisted 
of 200 ps of equilibration which was followed by 200 ps of data collection. 
The simulations were performed at NVT conditions where the temperature was kept constant 
at 298 K. Simulations of liquid methane and ethane were performed at
111.66 K and 184.52 K respectively. {As it was discussed in~\cite{prx}, there is no requirement 
of a coarse-grained model that resembles the structural and thermodynamic properties of a full atomistic
model. It was shown numerically that the proper exchange of energy and molecules was independent from the 
molecular model used in the coarse-grained region, showing the convenience of GC-AdResS. In this work, a generic WCA potential was used in the 
coarse-grained region. The interaction potential between the coarse-grained particles is given by 
\begin{equation}
  U(r) = 4 \epsilon \left[\left(\frac{\sigma}{r}\right)^{12} - \left(\frac{\sigma}{r}\right)^{6}\right] + \epsilon,
  \quad
  r \leq 2^{1/6}\sigma. 
\end{equation}
The parameters $\sigma$ and $\epsilon$ were chosen such that the radial distribution functions of particles 
reproduce a liquid structure. For water molecule, the parameters used in this study are $\epsilon=0.65$~kJ/mol
and $\sigma=0.30$~nm. Table~\ref{cg} shows the WCA parameters for other molecules used in this work.
For interactions between solute and solvent, $\sigma$ values were obtained by 
averaging over the individual parameters.
The solute-solvent $\epsilon$ is the same as the solute-solute  $\epsilon$.
\\
}

\begin{table}[]
\begin{center}
\begin{tabular}{ccc}
\hline \hline
 System & $\epsilon (kJ/mol)$ & $\sigma (nm)$ \\
\hline
methane & 0.65 & 0.40 \\
ethane & 0.20 & 0.50 \\
propane & 0.65 & 0.55 \\
methanol & 0.65 & 0.40 \\
DMSO & 0.30 & 0.50 \\
methanol in methanol/water & 0.65 & 0.40 \\
methane in methane/water & 0.65 & 0.40 \\
urea in urea/water & 0.65 & 0.40 \\
ethane in ethane/water & 0.65 & 0.45 \\
TBA in TBA/water & 0.65 & 0.60\\
DMSO in DMSO/water & 0.65 & 0.50 \\
TBA in TBA/DMSO & 0.40 & 0.60 \\
DMSO in TBA/DMSO & 0.30 & 0.50 \\
\hline \hline
\end{tabular}
\caption{WCA parameters for different coarse-grained molecules used in this work}
\label{cg}
\end{center}
\end{table}

To obtain the chemical potential of coarse-grained component, 
insertion particle method was used, where a trajectory of 8~ns was obtained and the coordinates were written
after every 0.4~ps. The insertions of the molecule were performed 4,000,000 times in each 
frame at random locations and with random orientations of the molecule.  
{The excess chemical potential value was calculated by averaging over the last ten iterations after 
the thermodynamic force has converged and the statistical uncertainty is determined by the standard deviation in the data. \\}

The excess chemical potential of the solute or the excess free energy of solvation was calculated using the 
thermodynamic integration (TI) approach. In the thermodynamic integration, the 
interaction of solute with the rest of the molecules in the systems is a function 
of a coupling parameter $\lambda$, which indicates a level of change taken place between states
A and B. The interactions are switched off as $\lambda$ is 
continuously decreased in the stepwise manner. Simulations conducted at different values of 
$\lambda$ allow to plot a {$\frac{\partial U_{i}(\lambda)}{\partial \lambda}$} curve, from which 
$\mu^{\exc}$ is derived \cite{mu}. 
{
\begin{equation}
 \mu^\exc_{iB} - \mu^\exc_{iA} = \int_{0}^{1} \left \langle \frac{\partial U_{i}(\lambda)}{\partial \lambda} \right \rangle_{\lambda} d\lambda 
\end{equation}
}
where $U_{i}$ is the interaction energy of particle $i$ with the remaining particles and $\langle \cdot \rangle$
denotes the canonical (NVT) or isobaric-isothermal (NPT) ensemble average. We computed the excess 
free energy using a two-stage approach as described in~\cite{mobley}, first coupling van der Waals interactions to transform the 
non-interacting molecule into a partially-interacting uncharged molecule, then coupling Coulomb
interactions from an uncharged interacting molecule to fully-interacting molecule. The resulting 
free energy $\Delta G_{final}$ is the sum of $\Delta G$ values obtained from the two procedures,
\begin{equation}
 \Delta G_{final} = \Delta G_{ele} + \Delta G_{vdw}
\end{equation}
where $\Delta G_{vdw}$ is the free energy change associated with introducing the van der Waals interactions and 
$\Delta G_{ele}$ is the free energy change associated with introducing Coulomb interactions. 
We evaluated the above integral for 21 values of $\lambda$ (evenly spaced between 0 and 1) in both the procedures.
At each value of $\lambda$, first a steepest descent energy minimization was performed followed 
by 200 ps of NPT equilibration and 400 ps of data collection under constant volume and temperature
conditions, in accordance with AdResS simulations. 
During the van der Waals coupling, soft-core interactions were used with soft-core parameters $\alpha_{LJ} = 0.5$, 
$\sigma = 0.3$ and the power of $\lambda$ in soft-core equation was taken as $1$. Free energy estimates and the errors
were calculated through Bennet's acceptance ratio method (BAR)~\cite{bar}. 
For both the AdResS and full-atom simulations, the system size was kept same.
{Table~\ref{table2} gives a detailed summary of each system studied}.\\

\begin{table}[]
\begin{center}
\begin{tabular}{ccccc}
\hline \hline
 System & $N_{solute}$ & $N_{solvent}$ & System size ($\textrm{nm}^{3}$) &  AT + HY region ($\textrm{nm}^{3}$)  \\
\hline
{water} & --- & 13824 & $30.2 \times 3.8 \times 3.8$ & $14.6 \times 3.8 \times 3.8$ \\
methane & --- & 2000 & $9.0 \times  3.7 \times 3.6$ & $6.0 \times 3.7 \times 3.6$ \\
ethane & --- & 2000 & $12.0 \times 3.9 \times 3.7$ & $7.0 \times 3.9 \times 3.7$ \\
propane & --- & 1433 & $10.0 \times 4.5 \times 4.5$ & $7.0 \times 4.5 \times 4.5$ \\
{methanol} & --- & 4000 & $12.0 \times 4.6 \times 4.5$ & $7.4 \times 4.6 \times 4.5$ \\
DMSO & --- & 1500 & $15.0 \times 3.6 \times 3.3$ &  $7.0 \times 3.6 \times 3.3$ \\
{methanol/water} & 128 & 12672 & $29.5 \times 3.7 \times 3.7$ & $14.6 \times 3.7 \times 3.7$ \\
{methane/water} & 40  & 6960 & $10.0 \times 4.8 \times 4.7$ & $7.0 \times 4.8 \times 4.7$ \\
urea/water & 50 & 2500 & $9.7 \times 2.9 \times 2.8$ & $6.8 \times 2.9 \times 2.8$ \\
{ethane/water} & 40 & 6960 & $10.0 \times 4.7 \times 4.6$ & $7.0 \times 4.7 \times 4.6$ \\
{TBA/water ($\concenttba=0.001$)} & 40 & 39960 & $50.1 \times 5.8 \times 4.3$ & $7.0 \times 5.8 \times 4.3$  \\
{TBA/water ($\concenttba=0.02$)} & 80 & 4400 & $10.0 \times 3.6 \times 4.2$ & $7.0 \times 3.6 \times 4.2$  \\
{TBA/water ($\concenttba=0.04$)} & 180 & 4300 & $10.0 \times 4.3 \times 3.7$ & $7.0 \times 4.3 \times 3.7$  \\
{TBA/water ($\concenttba=0.12$)} & 538 & 3942 & $10.0 \times 4.4 \times 4.6$ & $7.0 \times 4.4 \times 4.6$  \\
{TBA/water ($\concenttba=0.16$)} & 717 & 3763 & $10.0 \times 4.9 \times 4.6$ & $7.0 \times 4.9 \times 4.6$  \\
{TBA/water ($\concenttba=0.20$)} & 896 & 3584 & $12.0 \times 4.6 \times 4.5$ & $7.0 \times 4.6 \times 4.5$  \\
DMSO/water & 50 & 4950 & $12.0 \times 4.0 \times 3.3$ & $7.0 \times 4.0 \times 3.3$ \\
TBA/DMSO & 80 & 4400 & $10.0 \times 7.3 \times 7.2$ &  $7.0 \times 7.3 \times 7.2$ \\
\hline \hline
\end{tabular}
\caption{Summary of AdResS and full-atom systems.}
\label{table2}
\end{center}
\end{table}

In all the simulations, a leap-frog stochastic dynamics integrator with a time step
of 2~fs and an inverse friction coefficient of 0.1~ps was used. All bond-lengths were constrained using the LINCS 
algorithm.
For liquid water, methanol,
methanol/water, methane/water and ethane/water  a cut-off radius of
0.9 nm was used for van der Waals and Coulomb interactions, while for rest of the systems,
a cut-off radius of 1.4 nm was used.
{
  For the TBA/water system, the chemical potential converges at cut-off 0.9~nm for mole-fraction $\concenttba=0.02$.
  Since it would be too expensive to do the convergence tests for
  all concentrations, we simply use a large cut-off 1.4~nm for the concentration dependency study of TBA/water.}
Electrostatic interactions  were calculated using the reaction-field term \cite{rf} with a dielectric 
permittivity of 54 for urea in SPC water \cite{urea}, 64.8 for TBA in SPC water \cite{nico}, 61 for other solutes in SPC
water, 19 for methanol and 46 for DMSO as the solvent \cite{vang}.

\section{Technical Aspects of the Auxiliary Hamiltonian AdResS}\label{app:tmp2}
{In principle when the auxiliary Hamiltonian approach is used,
  one can perform microcanonical simulations and thus can avoid the
  use of a thermostat. In this case, the thermodynamic force of the
  auxiliary Hamiltonian would not carry any effect of the thermostat, and thus the difference between the work of the thermodynamic force of GC-AdResS and that of the auxiliary Hamiltonian is
  exactly the work that the thermostat does in GC-AdResS in order to compensate energy dissipation. The question is whether
  the energy is conserved in the auxiliary Hamiltonian approach. We
  have checked that the conservation holds for systems without
  electrostatics (methane,ethane,propane), thus for such systems the
  procedure is straightforward. Instead, for systems with
  electrostatic interactions, even for full atomistic simulations, due
  to the fact that the force fields are designed for employing the
  reaction field method, the energy cannot be conserved and the
  coupling to a thermostat is required. This is a well known problem
  reported in the manual of Gromacs. However, in our case, for both,
  the auxiliary Hamiltonian and GC-AdResS the energy drift due to the
  reaction field method is  essentially the same because they have
  equivalent electrostatic interactions, thus the energy drift due to
  the use of the reaction field method is automatically removed when
  we consider the difference between the thermodynamic forces of the
  two approaches, that is the force of changing resolution.}

\section{Extension of the chemical potential derivation to multi-component systems}\label{app:tmp3}

{In this section we  extend the chemical potential
expression of Eq.~\eqref{eqn:tmp24} to multi-component
systems, i.e. we show the derivation (and limitations) of Eq.~\eqref{eqn:tmp25}.
For simplicity and without lost of generality, we assume that the system
is formed by two components $\typea$ and $\typeb$, and the number of molecules are
$N^\typea$ and $N^\typeb$, respectively.  We further denote the number of 
molecules $\typea$ in the atomistic, transition, and coarse-grained regions by
$N^\typea_1$, $N^\typea_2$ and $N^\typea_3$, respectively and equivalently for type $\typeb$, $N^\typeb_1$, $N^\typeb_2$ and $N^\typeb_3$.  By
assuming, as usual, that the size of the transition region is negligible compared with the
atomistic and coarse-grained regions, we have the following
constrains:
\begin{align}
  V &= V_1 + V_3\\
  N &= N^\typea + N^\typeb\\
  N^\typea &= N^\typea_1 + N^\typea_3\\
  N^\typeb &= N^\typeb_1 + N^\typeb_3
\end{align}
We determine and apply the thermodynamic forces to each component, which
are denoted by $\vect F^\typea_\thf$ and $\vect F^\typeb_\thf$; thus we  impose the correct density profile to the system:
\begin{align}
  \rho_\HY^\typea &= \rho_\AT^\typea = \rho_\CG^\typea = \rho_0^\typea\\
  \rho_\HY^\typeb &= \rho_\AT^\typeb = \rho_\CG^\typeb = \rho_0^\typeb
\end{align}

Similarly to Eq.~\eqref{eqn:peq}, for pure systems, for a mixture in GC-AdResS we have:
\begin{align}\label{eqn:peq-d-ab}
  p_\CG - p_\AT=\rho_0^\typea \omega^\typea_\thf + \rho_0^\typeb \omega^\typeb_\thf
\end{align}
Following the same argument of Sec.~\ref{sec:tmp4}, we have
\begin{align}\label{eqn:peq-h-ab}
  p_\CG - p_\AT=
  \rho_0^\typea (\omega^{\typea,\hadress}_\thf - \omega^\typea_\res)
  +
  \rho_0^\typeb (\omega^{\typeb,\hadress}_\thf - \omega^\typeb_\res),
\end{align}
where the work of changing representation for molecule $\typea$ is defined by
\begin{align}\label{eqn:wrep-split-a}
  \omega^\typea_\res 
  &=
  \omega^{\typea\typea}_\res +\omega^{\typea\typeb}_\res
  = \int_\HY d\vect r \langle \vect F^{\typea\typea}_\res(\vect r)\rangle
  + \int_\HY d\vect r \langle \vect F^{\typea\typeb}_\res(\vect r)\rangle,\\
  \label{eqn:wrep-split-b}
  \omega^\typeb_\res 
  &=
  \omega^{\typeb\typea}_\res +\omega^{\typeb\typeb}_\res
  = \int_\HY d\vect r \langle \vect F^{\typeb\typea}_\res(\vect r)\rangle
  + \int_\HY d\vect r \langle \vect F^{\typeb\typeb}_\res(\vect r)\rangle.
\end{align}
Where ~$\omega_\res^{\typea\typea}$ denotes the work of changing representation for a molecule of $\typea$, due to the interaction with molecules of type $\typea$ only. Instead $\omega_\res^{\typea\typeb}$ denotes the work of changing representation for a molecule of $\typea$, due to the interaction with molecules of type $\typeb$ only. The same terminology holds for $\omega_\res^{\typeb\typea}$ and $\omega_\res^{\typeb\typeb}$.
The explicit expressions are:
\begin{align}
  \langle \vect F^{\typea\typea}_\res(\vect r)\rangle
  &=
  \nabla_{\vect r} w(\vect r)\,
  \big\langle w(\vect r')
  \big[\, \sum_{\alpha,\beta} U^\AT_{\typea\typea}(\vect r_\alpha- \vect r'_\beta)
  - U^\CG_{\typea\typea}(\vect r- \vect r') \,\big]
  \big\rangle_{\vect r',\typea; \vect r,\typea}\\
  \langle \vect F^{\typea\typeb}_\res(\vect r)\rangle
  &=
  \nabla_{\vect r} w(\vect r)\,
  \big\langle w(\vect r')
  \big[\, \sum_{\alpha,\beta} U^\AT_{\typea\typeb}(\vect r_\alpha- \vect r'_\beta)
  - U^\CG_{\typea\typeb}(\vect r- \vect r') \,\big]
  \big\rangle_{\vect r',\typeb; \vect r,\typea}\\
  \langle \vect F^{\typeb\typea}_\res(\vect r)\rangle
  &=
  \nabla_{\vect r} w(\vect r)\,
  \big\langle w(\vect r')
  \big[\, \sum_{\alpha,\beta} U^\AT_{\typeb\typea}(\vect r_\alpha- \vect r'_\beta)
  - U^\CG_{\typeb\typea}(\vect r- \vect r') \,\big]
  \big\rangle_{\vect r',\typea; \vect r,\typeb}\\
  \langle \vect F^{\typeb\typeb}_\res(\vect r)\rangle
  &=
  \nabla_{\vect r} w(\vect r)\,
  \big\langle w(\vect r')
  \big[\, \sum_{\alpha,\beta} U^\AT_{\typeb\typeb}(\vect r_\alpha- \vect r'_\beta)
  - U^\CG_{\typeb\typeb}(\vect r- \vect r') \,\big]
  \big\rangle_{\vect r',\typeb; \vect r,\typeb}
\end{align}
The notations are self-explanatory; for example $ U^\AT_{\typea\typeb}$
denotes the expression for atomistic interactions between one molecule of type $\typea$ and one of type $\typeb$ ($U^\AT_{\typea\typea}$ and$ U^\AT_{\typeb\typeb}$ are similar),
while $ U^\CG_{\typea\typeb}$ is the equivalent for coarse-grained interactions.
Notation $\langle\cdot\rangle_{\vect r',\typeb; \vect r,\typea}$ denotes the
ensemble average performed with respect to position $\vect r'$ of molecule
$\typeb$, provided that a molecule $\typea$ takes the position $\vect r$ (the same applies for other combinations on indices ${\vect r'}, {\vect r}, {\typea}, {\typeb}$).
If molecules contain more than one atom, then the average is also taken
over all possible conformations.
Therefore, the physical meaning of (for example) 
force $\langle \vect F^{\typea\typeb}_\res(\vect r)\rangle$
is that of an average force at $\vect r$ acting on a molecule of type $\typea$  due to the interaction with molecules of type $\typeb$.
Although we have $U^\AT_{\typea\typeb} = U^\AT_{\typeb\typea}$ and
$U^\CG_{\typea\typeb} = U^\CG_{\typeb\typea}$, it should
be noted that we do not have
$\omega_\res^{\typeb\typea} = \omega_\res^{\typea\typeb}$  in general.
From Eq.~\eqref{eqn:peq-d-ab}, \eqref{eqn:peq-h-ab},
\eqref{eqn:wrep-split-a} and \eqref{eqn:wrep-split-b}, we have
\begin{align}\label{eqn:hd-rel-ab}
  \rho_0^\typea(\omega_\thf^{\typea,\hadress} - \omega_\thf^\typea
  - \omega^{\typea\typea}_\res - \omega^{\typea\typeb}_\res)
  +
  \rho_0^\typeb(\omega_\thf^{\typeb,\hadress} - \omega_\thf^\typeb
  - \omega^{\typeb\typea}_\res - \omega^{\typeb\typeb}_\res) = 0
\end{align}
We denote the work done in the transition region on the two types
of molecules by $\omega_0^\typea$ and $\omega_0^\typeb$, respectively.
The chemical potential difference between the AT and CG resolution,
can be derived following the same procedure presented in Sec.III.C of Ref.5 which can be extended
to the two component system in a straightforward way.
Such a procedure leads to:
\begin{align}\label{eqn:tmpc7}
  \mu^\typea_\AT(N^\typea_1, N^\typeb_1, V_1, T) &= \mu^\typea_\CG(N^\typea_3, N^\typeb_3, V_3, T) - \omega^\typea_0\\\label{eqn:tmpc8}
  \mu^\typeb_\AT(N^\typea_1, N^\typeb_1, V_1, T) &= \mu^\typeb_\CG(N^\typea_3, N^\typeb_3, V_3, T) - \omega^\typeb_0
\end{align}
In the thermodynamic limit, these numbers maximize the Helmholtz free energy.
In this context the chemical potential, e.g.~$\mu_\AT^\typea$, is the free energy increment due to the insertion of one molecule of type $\typea$ 
into the infinitely large $\typea$--$\typeb$ mixture.\\
Similarly to the case of the one component system, from Eq.~\eqref{eqn:tmpc7} and \eqref{eqn:tmpc8},
we write down  for GC-AdResS:
\begin{align}\label{eqn:mueq-d-a}
 \mu^\typea_\CG - \mu^\typea_\AT &=   \omega^\typea_\thf + \omega^\typea_\dof + \omega^\typea_\ext\\\label{eqn:mueq-d-b}
 \mu^\typeb_\CG - \mu^\typeb_\AT &=   \omega^\typeb_\thf + \omega^\typeb_\dof + \omega^\typeb_\ext
\end{align}
$ \omega^\typea_\thf$ and $ \omega^\typeb_\thf$ are 
the work of the thermodynamic force $\vect F^\typea_\thf$ and
$\vect F^\typeb_\thf$, respectively. $\omega^\typea_\ext$ is the energy
dissipation due to molecule $\typea$ that
changes resolution in the transition region,
and  $\omega^\typeb_\ext$ is defined similarly.
The energy dissipation can be further divided as:
\begin{align}\label{eqn:wextra-split-a}
  \omega^\typea_\ext& = \omega^{\typea\typea}_\ext + \omega^{\typea\typeb}_\ext\\\label{eqn:wextra-split-b}
  \omega^\typeb_\ext& = \omega^{\typeb\typea}_\ext + \omega^{\typeb\typeb}_\ext
\end{align}
$ \omega^{\typea\typea}_\ext$ is the energy dissipation of a molecule
$\typea$ produced by  non-conservative interactions between
molecule type $\typea$ and type $\typea$ only.
Similarly $\omega^{\typea\typeb}_\ext$ is
the energy dissipation of a molecule $\typea$
due to the non-conservative interactions with molecules of type $\typeb$.  The definitions are similar for $
\omega^{\typeb\typea}_\ext$ and $ \omega^{\typeb\typeb}_\ext$.
It should be noticed that, we do not have
$\omega^{\typeb\typea}_\ext = \omega^{\typea\typeb}_\ext$ in general.
For the expression of the chemical potential, the same argument as above, is applied to the auxiliary Hamiltonian approach, and
yields
\begin{align}\label{eqn:mueq-h-a}
 \mu^\typea_\CG - \mu^\typea_\AT &=   \omega^{\typea,\hadress}_\thf + \omega^\typea_\dof \\\label{eqn:mueq-h-b}
 \mu^\typeb_\CG - \mu^\typeb_\AT &=   \omega^{\typeb,\hadress}_\thf + \omega^\typeb_\dof
\end{align}
By using Eq.~\eqref{eqn:mueq-d-a}, \eqref{eqn:wextra-split-a} and \eqref{eqn:mueq-h-a}, we have
\begin{align}\label{eqn:hd-rel-2-a}
  \omega^{\typea\typea}_\ext + \omega^{\typea\typeb}_\ext
  = \omega^{\typea,\hadress}_\thf - \omega^\typea_\thf.
\end{align}
Using Eq.~\eqref{eqn:mueq-d-b}, \eqref{eqn:wextra-split-b} and \eqref{eqn:mueq-h-b}, we have
\begin{align}\label{eqn:hd-rel-2-b}
  \omega^{\typeb\typea}_\ext + \omega^{\typeb\typeb}_\ext
  = \omega^{\typeb,\hadress}_\thf - \omega^\typeb_\thf.
\end{align}

By inserting Eq.~\eqref{eqn:hd-rel-2-a} and \eqref{eqn:hd-rel-2-b} into
Eq.~\eqref{eqn:hd-rel-ab}, we have
\begin{align}\label{eqn:hd-rel-ab-1}
  \rho_0^\typea(\omega_\ext^{\typea\typea} + \omega_\ext^{\typea\typeb}
  - \omega^{\typea\typea}_\res - \omega^{\typea\typeb}_\res)
  +
  \rho_0^\typeb(\omega_\ext^{\typeb\typea} + \omega_\ext^{\typeb\typeb}
  - \omega^{\typeb\typea}_\res - \omega^{\typeb\typeb}_\res) = 0
\end{align}
It is natural to conclude that
$\omega_\ext^{\typea\typea} = \omega_\res^{\typea\typea} $,
because  these two terms exclusively
involves $\typea$--$\typea$ interaction.
The same is true for $\typeb$--$\typeb$ interaction:
$\omega_\ext^{\typeb\typeb} = \omega_\res^{\typeb\typeb} $.
The physical meaning of $\omega_\ext^{\typea\typeb}$,
$\omega_\res^{\typea\typeb}$, $\omega_\ext^{\typeb\typea}$ and
$\omega_\res^{\typeb\typea}$, leads to 
identify of $\omega_\ext^{\typea\typeb}$ with $\omega_\res^{\typea\typeb}$,
and $\omega_\ext^{\typeb\typea}$ with $\omega_\res^{\typeb\typea}$.
It follows that (for example) for component $\typea$, the excess chemical potential difference is:
\begin{equation}
\mu^{\typea,\exc}_{\CG}-\mu^{\typea,\exc}_{\AT}=\int _{\Delta}{\vect F}^{\typea}_{\thf}({\vect r})d{\vect r}+\int_{\Delta} \langle \vect F^{\typea\typea}_\res(\vect r)\rangle d{\vect r}+\int_{\Delta} \langle \vect F^{\typea\typeb}_\res(\vect r)\rangle d{\vect r}
\label{fineq}
\end{equation}
 and this proves Eq.~\eqref{eqn:tmp25}.}

\end{document}